\documentstyle[12pt]{article}
\parskip=\medskipamount	
\newcommand{\be}{\begin{equation}}
\newcommand{\ee}{\end{equation}}
\newcommand{\bea}{\begin{eqnarray}}
\newcommand{\eea}{\end{eqnarray}}

\let\newsection=\section
\renewcommand{\section}{\setcounter{equation}{0}\newsection}

\def\on#1#2{{\buildrel{\mkern2.5mu#1\mkern-2.5mu}\over{#2}}}
\def\dt#1{\on{\hbox{\bf .}}{#1}}                
\def\tdt#1{\on{\hbox{\bf .\kern-1pt.\kern-1pt.}}#1}   
\def\half{{\textstyle{1\over{\raise.1ex\hbox{$\scriptstyle{2}$}}}}}
\def\bop#1{\setbox0=\hbox{$#1M$}\mkern1.5mu
	\vbox{\hrule height0pt depth.04\ht0
	\hbox{\vrule width.04\ht0 height.9\ht0 \kern.9\ht0
	\vrule width.04\ht0}\hrule height.04\ht0}\mkern1.5mu}
\def\Box{{\mathpalette\bop{}}}                        

\begin{document}
\begin{flushright}
YITP-SB-00-57\\
IASSNS-HEP-00/80\\\
hep-th/0010106\\
\end{flushright}
\vskip.5in

\begin{center}

{\LARGE\bf A new AdS/CFT correspondence}\\[.5in minus.2in]

{\bf H. Nastase}\\[.1in]
{\it C.N. Yang Institute for Theoretical Physics\\
	State University of New York, Stony Brook, NY 11794-3840}\\[.1in]
and \enskip{\it School of Natural Sciences\\
	Institute for Advanced Study, Princeton, NJ 08540}\\[.1in]
e-mail:  {\tt nastase@ias.edu}\\[.3in]
{\bf W. Siegel}\\[.1in]
{\it C.N. Yang Institute for Theoretical Physics\\
	State University of New York, Stony Brook, NY 11794-3840}\\[.1in]
e-mail:  {\tt siegel@insti.physics.sunysb.edu}

\end{center}
\vskip1in minus.8in

\begin{abstract}

We consider a geometric zero-radius limit for strings on $AdS_5\times
S^5$, where the anti-de Sitter hyperboloid becomes the projective
lightcone.  In this limit the fifth dimension becomes nondynamical,
yielding a different ``holographic" interpretation than the usual ``bulk
to boundary" one.  When quantized on the random lattice, the fifth
coordinate acts as a new kind of Schwinger parameter, producing
Feynman rules with normal propagators at the tree level:  For example, in
the bosonic case ordinary massless $\phi^4$ theory is obtained.  In the
superstring case we obtain new, manifestly ${\cal N}=4$ supersymmetric
rules for ${\cal N}=4$ super Yang-Mills.  These gluons are also different
from those of the usual AdS/CFT  correspondence:  They are the
``partons" that make up the usual ``hadrons" of the open and closed
strings in the familiar QCD string picture.  Thus, their coupling $g_{YM}$
and rank $N$ of the ``color" gauge group are different from those of the
``flavor" gauge group of the open string.  As a result we obtain different
perturbation expansions in radius, coupling, and $1/N$.

\end{abstract}
\newpage

\section{Outline}

The AdS/CFT correspondence \cite{1} derives a four-dimensional
conformal field theory from Type IIB superstring theory by choosing as a
vacuum five-dimensional anti-de Sitter space, whose symmetry SO(4,2) is
the same as the four-dimensional conformal group.  The Maldacena
conjecture is that all the important dynamics (at least in an appropriate
limit) takes place on the four-dimensional boundary of that space
(``holography").  The remaining five dimensions of the superstring, which
form a sphere, contribute the SO(6) internal symmetry of the ${\cal N}=4$
superconformal group:  Including the fermions, this superstring then
describes the color singlets of ${\cal N}=4$ super Yang-Mills theory.

In this paper we will find a similar correspondence (``duality") between
this same string theory and a different maximally supersymmetric
Yang-Mills theory.  In the following section we describe our different
holography:  It is based on a zero-radius limit where $AdS_5$ shrinks at
the center to become a (projective) lightcone.  It also emphasizes the
boundary, but the fifth dimension remains as a nondynamical (to leading
order) auxiliary variable.  The construction is illustrated by the case of
the particle.  (It is a geometric construction that does not require the
explicit appearance of a string.)  In section 3 we extend this limit to an
expansion.  It uses the same couplings available in the Maldacena
description, but after a rescaling of the coordinates.

The other method we apply is the random worldsheet lattice, introduced
in section 4.  This approach focuses on fields that do not appear explicitly
otherwise:  Upon identifying the discretized path integral as Feynman
diagrams of a matrix field theory, these ``constituent" fields of the
``composite" string states are naturally identified in a QCD string picture,
where ${\cal N}=4$ supersymmetric Yang-Mills is a ``simpler" analog of
QCD.  Their gauge group and coupling are not the same as those of the
similar ground states of the open string:  In QCD language, ``gluons" are
not the same as massless (in this idealized model) ``$\rho$ mesons".

These two methods are combined in section 5 in application to the
bosonic string, which illustrates how the fifth dimension reduces, in
leading order, to a kind of Schwinger parameter.  The result is that the
tree graphs of this model are identical to those of massless
(wrong-sign) $\phi^4$ theory (in contrast to random lattice quantization
about flat space, which produces Gaussian propagators).  In section 6 this
derivation is used to relate the couplings of the closed string, open string
(a la Maldacena), and random matrix theory.

In the final section we extend the analysis to the superstring.  The limit
kills the Wess-Zumino term (as well as shrinking the sphere $S^5$ to a
point).  In a (probably unrealistic, but covariant) ``long string" gauge, the
tree graph Feynman rules are those that follow from a manifestly ${\cal
N}=4$ supersymmetric action for a matrix field that we identify with the
super Yang-Mills field.

\section{Holographies}

The use of higher dimensions to manifest conformal symmetry was
originally proposed by Dirac \cite{2,3} in a form different from AdS/CFT:  In
six dimensions, with appropriate signature, the Lorentz group alone
yields the four-dimensional conformal group.  Restriction to the lightcone
eliminates one coordinate, preserving this group (but breaking the
unwanted 6D translational invariance).  This 5D lightcone is a limit of 5D
anti-de Sitter space:  In terms of 6D coordinates $z$, the latter is
described by
$z^2+R^2=0$ for some radius $R$ (giving constant 5D curvature $-1/R^2$),
and the former by $z^2=0$.  The fifth coordinate can then be eliminated
by introducing a scale invariance, identifying $z$ with any real multiple of
itself; fields on this space are required to be homogeneous in $z$.  The
resulting projective lightcone yields a description of conformal field
theories equivalent to the usual 4D one, but makes conformal symmetry
manifest.  (A version of this approach was the basis of the ADHM
construction of instantons \cite{4}.)

Our approach will be to perturb the string action in $R$:  The lowest-order
approximation to anti-de Sitter space will then be the projective
lightcone.  Interpreting holography in the more general sense, as
describing dynamics with one less spatial dimension, we therefore have a
new holographic approach to AdS/CFT (although this difference alone will
not yet give our new AdS/CFT correspondence).  As for the Maldacena
holography, this is not a proof of holography, but rather an expansion
about a holographic limit.

In practice, the Maldacena approach is usually applied by first reducing
the string to 5D gauged supergravity, i.e., truncating the 10D superstring
to a 5D superparticle.  For an explicit comparison of the two holographies,
we first consider the simpler bosonic analog, the classical mechanics of a
conformal scalar particle in a background scalar field.  Upon
(first-)quantization, this approach relates directly to pure field theory, as
a particular set of Feynman tree graphs for $\phi^n$ theory can be
represented in terms of the propagator of a particle in a $\phi^{n-2}$
background, simply modifying $p^2\to p^2+\phi^{n-2}$.

A particle in anti-de Sitter space $AdS_{D+1}$ for arbitrary dimensions $D$
is then described by
\be
S = \int d\tau\; [ \half\dt z{}^2 +\half\lambda(z^2+R^2) +\hat\xi(z) ]
\ee
  with Lagrange multiplier $\lambda$ and (perhaps composite) background
scalar field $\hat\xi$.  The relation to the usual 4D coordinates is given
by
\be
x^a = {z^a\over z^+}, \quad x^0 = {1\over z^+}
\ee
  in a lightcone basis $(z^+,z^-,z^a)$ with respect to the two extra
coordinates.  (We also identify $z\to -z$ by restricting $x^0\ge 0$.)  The
restriction to the boundary
\be
\hat\xi(z) = \delta(x^0)\xi(x^a)
\ee
  is applied as a limit.  Solving the constraint on $z$ from varying
$\lambda$,  the action then reduces to
\be
S = \int d\tau\; \left[ \half{(\dt x{}^a)^2 +R^2(\dt x{}^0)^2 \over
	(x^0)^2} +\delta(x^0)\xi(x^a) \right]
\ee
  and $x^0$ remains dynamical (although we can decouple it in the limit
$R\to 0$).  The result is not the usual 4D scalar.

On the projective lightcone, the Lagrangian form of the action is the
result of taking $R\to 0$ on the previous:
\be
S = \int d\tau\; [ \half\dt z{}^2 +\half\lambda z^2 +\hat\xi(z) ]
\ee
  Projective invariance is incorporated into reparametrization invariance
\cite{5}:
\be
\delta z = \epsilon\dt z -\half\dt\epsilon z, \quad
\delta\lambda = \epsilon\dt\lambda
+2\dt\epsilon\lambda-\half\tdt\epsilon
\ee
  This is preserved by the background only if
\be
z\cdot{\partial\over\partial z}\;\hat\xi = -2\hat\xi
\ee
   This homogeneity condition on the background is solved as
\be
\hat\xi(z) = (x^0)^2\xi(x^a)
\ee
  Again solving the constraint, the action becomes the usual
\be
S = \int d\tau\; [ \half (x^0)^{-2}(\dt x{}^a)^2 +(x^0)^2\xi(x^a) ]
\ee
  where $x^0$ has not disappeared, but has become the worldline metric.
Although nondynamical, it plays an important role upon quantization:  It
yields the Schwinger parameters.  The result is the standard description
of a real massless scalar in a background.  The background can itself be
identified with the usual bosonic field by dimensional analysis:  Since the
background is the usual conformal scalar field theory interaction term
less two powers of the scalar, we have
\be
  \hat\xi = \hat\phi^{4/(D-2)} \quad\rightarrow\quad
  \hat\phi(z) = (x^0)^{(D-2)/2}\phi(x^a)
\ee
  i.e., $\hat\xi=\hat\phi^2$ and $\hat\phi=x^0\phi$ in $D=4$.

\section{Expansions}

There are two dimensionless couplings in the $AdS_5\times S^5$
description of Type IIB superstring theory (after including appropriate
powers of $\alpha'$).  One is the ``string coupling", which appears for any
background (``vacuum"), and is associated with the asymptotic value of
the dilaton field.  It is included in the worldsheet action through a
topological term, which we will include implicitly.  Both of the expansions
we will consider expand in this coupling.

The other is the radius $R$ of $AdS_5$ and $S^5$.  We have already
included it, but it can be moved around through coordinate
transformations, which are always allowed in a gravitational theory (i.e.,
the string).  (However, we will only need to redefine $x^0$, which is not
considered ``observable", so the worst consequence will be field
renormalizations.)  In the Feynman rules that we will derive, such
rescalings of coordinates are equivalent to rescalings of fields, since
propagators (and in general, vertices) in conformal theories are powers
of coordinates (or momenta); it is the equivalence between active and
passive transformations.  However, the coupling expansions are
dependent on such redefinitions:  For convenience, we always expand in
$R$ or $1/R$ while keeping $x^a$ and $x^0$ fixed, so different expansions
can be obtained by $R$-rescalings of the coordinates.

In particular, the $AdS$ metric we used above, motivated by geometrical
considerations, is
\be
  ds^2 = {(dx{}^a)^2 +R^2(dx{}^0)^2 \over (x^0)^2}
\ee
  It differs from that useful in the Maldacena approach,
\be
  ds^2 = R^2{(dx'{}^a)^2 +(dx'{}^0)^2 \over (x'^0)^2}
\ee
  through a rescaling of $x^0$,
\be
  x^0 = {x'^0\over R}
\ee
  or $x^a=Rx'^a$.  Specifically, we expand $ds^2$ in $R^2$, $x$ fixed,
while the other approach expands in $1/R^2$, $x'$ fixed.  The former is a
geometrical expansion of $AdS$ about the (projective) lightcone, while
the latter is a JWKB expansion, since $R^2$ appears as an overall factor in
the string mechanics action.  In the Maldacena coordinates, our expansion
corresponds to the limit $R\to 0$, $x'^0\to 0$, $x'^0/R$ fixed:  It is an
expansion about the boundary, so the two holographies are closely
related.  However, in our holography $x^0$ remains as a coordinate, even
after taking the limit (i.e., for the leading term in the expansion):  As we
saw in the example of the previous section, the limiting theory is not
expressed directly as a boundary theory, but as a theory where one
spatial dimension is nondynamical, although still playing a useful role.

Note that when we say ``Maldacena approach" we begin with
$AdS_5\times S^5$.  Maldacena actually began with a different,
asymptotically flat space, where the relationship to D-branes (and thus
the open string Yang-Mills group) was clear, and took an additional limit
to obtain $AdS_5\times S^5$.  For us there is no advantage to starting
with the other space.  Both are 3-brane solutions to the classical
equations, differing by a constant of integration, where asymptotically
flat space effectively has an extra 3-brane at the boundary:  Explicitly,
solutions for the various background fields (including the metric) with
parallel BPS (small supersymmetry multiplet) 3-branes are
characteristically expressed in terms of a function of the form
\be
a +\sum_i {b_i^2\over (y-y_i)^4}
\ee
  for some constants $a,b_i,y_i$, where $y$ are the coordinates
``orthogonal" to the branes:  $y_i$ are their positions, and $b_i$ their
charges.  The constant of integration $a$ is generally fixed to 1 for an
asymptotically flat boundary, but otherwise is arbitrary, and can also be
set to 0.  In particular, it can be generated by choosing one (or some) of
the branes to be at $y_1=\infty$ (with $b_1=\infty$).  In the Maldacena
case, the remaining $y_i$'s are set to 0.  We are thus free to set $a=0$
from the beginning as a choice of background solution, rather than
starting with a less desirable choice and requiring an additional,
irrelevant limit.  The choice $a=0$ is difficult to interpret in a
conventional S-matrix approach (although not an obstruction in either a
$\beta$-function or string field theory approach); but it is exactly the
unconventional properties (even the definition) of the S-matrix in
(super)conformal field theories that is being studied in the AdS/CFT
correspondence.

So far we have excluded the $\alpha'$ dependence of the action:  After
the same kind of redefinitions we have made for $R$, it always appears in
the combination $R^2/\alpha'$ in $AdS_5\times S^5$.  In any case, it can
always be restored by the substitution $R^2\to R^2/\alpha'$.  (In the
particle case any such overall factor in the spacetime metric can be
absorbed into the worldline metric because of lack of worldline scale
invariance.)  We will thus always refer to $R^2$ as either the radius
(squared) or the tension (although it is a little more like a torque).
This interpretation is analogous to that of $\hbar$ and the coupling $g$ in
any gauge field theory:  There the dimensionless combination is $\hbar
g^2$, and the two appear only in this combination, after appropriate field
redefinitions.  Expansion in this $\hbar$ produces the usual loop expansion
of field theory.  However, we can redefine $x\to x/\hbar$ and $\hbar
g^2\to g^2/\hbar$; then expansion in $\hbar$ produces the JWKB
expansion of (relativistic) quantum mechanics \cite{6}.  Introducing
$\hbar$ into field theory defines the dimension of ``mass" independently
from ``length", but necessarily in a trivial way if the theory is scale
invariant.
Thus, we could write
\be
  ds^2 = {1\over\alpha'}{(dx{}^a)^2 +R'^2(dx{}^0)^2 \over (x^0)^2}, \qquad
R^2 \equiv {R'^2\over \alpha'}
\ee
  Then the Maldacena interpretation comes from setting $R'=1$ and
expanding in $\alpha'$, while our interpretation sets $\alpha'=1$ and
expands in $R'$.  In either case $R$ remains as a nontrivial dimensionless
coupling.

\section{Random lattices}

Besides the $R$ expansion, which leads to the projective lightcone, the
other main ingredient in our analysis is the random worldsheet lattice
\cite{7}. Although either of these two methods can be applied separately
to the string on $AdS_5\times S^5$, we will find that the combination
solves some mutual problems.  The random lattice approach is based on
introducing a lattice regularization for the worldsheet coordinates.  Since
the worldsheet is curved, the lattice is irregular, and the sum over
geometries in the path integral becomes a sum over lattices of different
geometries.  By identifying the lattices as Feynman diagrams, the
first-quantized path integral for the string becomes a sum of Feynman
diagrams for some {\it particle} field theory, whose propagators and
vertices are read from the discretized action of the string.  This
discretization is used in conjunction \cite{8} with the $1/N$ expansion
\cite{9}, which allows a 2D topology to be associated with any Feynman
diagram by using fields that are $N\times N$ matrices.  String states are
identified with singlets under the corresponding $U(N)$ symmetry:  In a
QCD interpretation, the Feynman rules are for a chromodynamic field
theory, describing gluons and quarks, while the string describes hadrons.
The motivation for such a dual approach to strong interactions is that one
could continue to use perturbative QCD to describe processes with large
transverse momenta while applying string theory to describe the
asymptotic spectrum and Regge behavior: two different practical
perturbation expansions for the same theory, each accurate in its region
of momentum space.

Thus the random lattice approach leads to a different interpretation of
the various Yang-Mills fields than that used in the Maldacena approach.
In the hadronic interpretation of string theory, open strings describe
mesons, while closed strings describe ``pomerons".  In QCD strings and
their generalizations (super Yang-Mills strings, etc.), the mesons are
identified as quark-antiquark states bound by gluons, while the pomerons
are identified as ``glueballs".  The random lattice approach explicitly
associates a QCD-like particle theory of ``partons" (gluons and quarks)
with a hadron-like string theory of mesons and pomerons.  Previously
little explicit success has been obtained with this approach except for the
bosonic string, using scalar partons.  Here we propose, through an
AdS/CFT approach, to identify maximally supersymmetric Yang-Mills
theory with the partons, as the only known maximally superconformal
theory with fields than can be identified as matrices with respect to an
internal symmetry.  Of course, in this (unbroken) supersymmetric model,
fermions come in unphysical group representations, and the lightest
hadrons appear as massless (as in the original Maldacena model).  Thus,
while the partons include massless gluons, the open string states include
massless $\rho$ mesons.  Both the gluons and the $\rho$'s appear in
supersymmetric multiplets, and both are associated with gauge fields.
(Originally, Yang and Mills identified nonabelian gauge fields as massless
$\rho$'s.  After the Higgs effect was discovered, but before QCD,
phenomenological models of low-mass hadrons made the $\rho$'s
massive by eating the scalars of linear or nonlinear $\sigma$-models.)
This differs from the Maldacena interpretation, where the massless
open-string states are themselves identified with gluons, while the
closed string states are still identified with pomerons (glueballs).  Our
interpretation thus more closely coincides with the original QCD
interpretation, where the gluons are not themselves strings.  (However,
as an alternative interpretation of our approach, the states of the
underlying particle theory could be identified instead as ``preons".)
Independent of interpretation, the random lattice approach necessarily
involves two different groups, as the group acting on the matrices of the
underlying particle theory is not necessarily identical to the group acting
on the open string states.  More importantly, our AdS/CFT correspondence
necessarily differs from that of Maldacena because the conformal field
theory we analyze is not that of the open string states, but that of the
underlying particle theory.  As a consequence, unlike the Maldacena
approach, in our AdS/CFT correspondence the Yang-Mills fields that
we identify with the gluons appear explicitly, so in principle both
perturbative and nonperturbative factors of QCD amplitudes can be
examined together.  Furthermore, in the D-brane picture of the open
strings, where a closed string can create an open string pair (as in Type I
theory, except here for Type II the ends of the open strings are confined
to the D3 branes), the random lattice treats both the open and closed
strings as flux tubes of the partonic gluons:  As in the 't Hooft picture for
QCD, a closed string (pomeron/glueball) can still be considered a bound
state of two open strings (mesons), but both types of strings are
coherent states of arbitrary numbers of gluons; the glueballs are really
closed flux tubes, while the open strings are open flux tubes.

One problem with the random lattice approach is that quadratic string
actions give Gaussian propagators:  Discretizing the usual $(\partial x)^2$
term common to all strings,
\be
  \int Dx\; e^{-S} \sim \int\left(\prod_I dx_I\right)
  e^{-\half\sum_{\langle IJ\rangle} (x_I-x_J)^2}
\ee
  for vertices $I$ with links (propagators) $\langle IJ\rangle$.  (Other
terms in superstring actions yield vertex factors, and ``numerators" for
the propagators \cite{10}.)  Consequences are non-parton-like behavior at
large transverse momenta \cite{11}, and essentially no degrees of
freedom past the Hagedorn temperature \cite{12} (since the propagator
has no poles), in what should be the parton plasma phase.  One proposal
was to introduce Schwinger parameters as a new degree of freedom in
the string action \cite{13}, but no correspondence with the usual strings
was obtained.  However, the example of section 1 suggests a more
natural way to incorporate the extra variable within the conventional
string framework. (The propagators in the Maldacena approach are not
the usual ones, nor Gaussians, but require Bessel functions.)

\section{Bosonic string}

The analysis of section 1 generalizes straightforwardly (except for
reparametrizations of the worldline/sheet) to (five dimensions of) the
bosonic string:  Since we use the same background spacetime metric, we
need only replace derivatives and integrals with worldsheet ones,
$d\tau\to d\tau d\sigma$, $\dt f{}^2\to (\partial f)^2$, where the former
implicitly includes the worldsheet measure $\sqrt{-g}$, and the latter the
worldsheet metric as $g^{mn}(\partial_m f)(\partial_n f)$.  (We consider
here $AdS_5$ and ignore effects of worldsheet and spacetime conformal
anomalies.)  Then introducing the random lattice as in the previous
section, the action
\bea
S &=& \int d^2\sigma\; \sqrt{-g}[\half g^{mn}(\partial_m z)(\partial_n z)
  +\half\lambda(z^2 +R^2)] \nonumber \\[.1in]
&\to& \sum_{\langle IJ\rangle}\half(z_I-z_J)^2
  +\sum_I\half\lambda_I(z_I^2+R^2)
\eea
  is path integrated, with insertions of background sources (``punctures")
$\zeta(z)$, as
\be
{\cal A} = \int Dz D\lambda\; e^{-S}\zeta ... \zeta \quad\to\quad
  \int\left(\prod_I dz_I d\lambda_I\right) e^{-S}\zeta ... \zeta
\ee
  Integrating out $\lambda$,
\be
{\cal A} = \int\left[\prod_I dz_I\delta(z_I^2+R^2)\right]
  e^{-\half\sum_{\langle IJ\rangle} (z_I-z_J)^2}
  \prod_{I'}\zeta(z_{I'})
\ee
  where $I'$ are just the points where the background is attached.  (For
simplicity, we have introduced pointlike sources, corresponding to ground
states, e.g., $e^{ik\cdot x}$.  Sources for $n$th derivatives use functions
of $n$ adjacent points, as $\partial x\to x_I-x_J$ in the action.)

Note that we have not determined the $R$ dependence of the measure; it
can easily be fixed at the end, as we will describe in the following
section.  We have also used the simplest discretization of the action:  It
actually corresponds not to the true distance squared, but the distance as
would be measured in the six-dimensional embedding space.  Lattice
actions are always ambiguous to such ``higher-derivative" terms; our
choice is not only simpler than the true distance, but single-valued at the
antipode in the case of the sphere.  (In that case, we have replaced the
distance $s$ with $2\,sin(s/2)$.)

In the random lattice approach, 't Hooft's analysis of the $1/N$ expansion
is used to identify 2D topology of Feynman diagrams:  The double lines
representing the gluon propagators give the gluons a slight
``stringiness", so that the leading order in $1/N$ is identified pictorially
with a diagram that is planar with respect to the double lines.  The
double lines are a consequence of the use of a matrix representation,
and can be applied to any theory whose fields are matrices.  The value of
the Euler number $\chi$ of an arbitrary diagram can be identified
unambiguously, and contributes a factor $N^\chi$, with all remaining
$N$-dependence occuring together with the defining-representation
coupling $g^2$ (appropriate to the double-line notation) only in the
combination of the adjoint-representation coupling $Ng^2$.  Since explicit
color lines do not appear until the introduction of the double lines, the
sources appearing in the path integral above are necessarily color
singlets. However, the Feynman rules for color nonsinglets, and in
particular the matrix fields themselves, are obvious from the rules for
singlets.

Again identifying the discretized first-quantized string path integral as a
Feynman diagram, and introducing the double lines, we read off the
Feynman rules as
\bea
\hbox{Propagator:} && \delta_{i'}^j\delta_i^{j'}e^{-(z-z')^2/2}
	\nonumber \\
\hbox{Vertex:} && \delta_{j_2}^{i_1}\delta_{j_3}^{i_2}...\delta_{j_1}^{i_n}
	\textstyle{\int} d^6 z\;\delta(z^2+R^2) \nonumber \\
\hbox{External line:} && \hat\phi_i{}^j(z)
\eea
  This external line is for a background matrix field $\hat\phi$ itself; the
color singlet sources considered above are traces of powers of such fields.
We have associated the worldsheet with Feynman diagrams with
$n$-point vertices; (spacetime) conformal invariance will restrict this
choice.

  To compare with the usual 4D rules, we replace $z\to(x,x^0)$ as before.
We find
\be
  (z-z')^2 = {(x-x')^2 +R^2(x^0-x'^0)^2\over x^0 x'^0}
\ee
For the projective lightcone limit $R\to 0$, we also replace
$\hat\phi=x^0\phi(x)$.  The rules then become, now dropping indices,
\bea
\hbox{Propagator:} && e^{-(x-x')^2/2x^0 x'^0} \nonumber \\
\hbox{Vertex:} && \textstyle{\int} d^4 x\;dx^0(x^0)^{-5} \nonumber \\
\hbox{External line:} && x^0\phi(x)
\eea

Because of the simple $x^0$ dependence in the projective lightcone limit,
this coordinate can easily be integrated out explicitly in tree graphs.  The
easiest way is to examine the Schwinger-Dyson equation (what would be
the integrated field equation):  Choosing as random lattices the diagrams
of
$\phi^4$ theory (i.e., 4-point vertices only),
\be
\hat\phi(z) = \int d^6 z'\;\delta(z'^2)e^{-(z-z')^2/2}\hat\phi^3(z')
\ee
  which is now
\be
x^0\phi(x) = \int d^4 x'dx'^0(x'^0)^{-5}e^{-(x-x')^2/2x^0 x'^0}(x'^0)^3
  \phi^3(x')
\ee
  Integrating $\int_0^\infty dx'^0$,
\be
\phi(x) = \int d^4 x'{1\over\half(x-x')^2}\phi^3(x')
\ee
  This is the usual Schwinger-Dyson equation for massless $\phi^4$ theory,
as $x^{-2}$ is the usual massless propagator.  (Of course, the integrals
can be evaluated as easily in momentum space, or for conformal scalars
in other dimensions.  If we like, we can also add a source term as
$\hat\phi^3\to\hat\phi^3+\hat J$, or $\phi^3\to\phi^3+J$, as
$\hat J=(x^0)^3 J(x)$.)  Thus the role of $x^0$ in the projective lightcone
expansion has been reduced effectively to that of Schwinger parameters.
(Extra dimensions act similarly to Schwinger and Feynman parameters in
the Parisi-Sourlas formalism also \cite{14}, particularly when applied to
covariantizing the usual lightcone \cite{15}.)

The final Feynman rules for tree graphs are therefore the usual ones of
massless
$\phi^4$ theory:
\bea
\hbox{Propagator:} && 1/\half (x-x')^2 \nonumber \\
\hbox{Vertex:} && \textstyle{\int} d^4 x \nonumber \\
\hbox{External line:} && \phi(x)
\eea
  (Adler \cite{3} also formulated Feynman rules for conformal theories on
the projective lightcone, but he set $x^0=1$ by hand as a gauge condition,
which is not applicable here.)  Note that the usual massless propagator
resulted only because we assumed the usual conformal $\phi^4$
interaction:  If we had assumed a different power of $\hat\phi$ (with
$x^0$ dependence consistent with conformal invariance), a different
power of $x^2$ would have appeared, consistent with the modified scale
weight of $\phi$, but implying a nonlocal kinetic operator.

We can see the above procedure implemented explicitly in any tree graph,
integrating one vertex at a time.  Unfortunately, the final integration in
any diagram will always produce an extra factor
\be
\int_0^\infty {dx^0\over x^0} = \infty
\ee
  This factor appears with each connected graph; it accompanies the usual
energy-momentum conservation $\delta$-function
\be
\delta\left(\sum p\right) \sim \int d^4 x = \infty
\ee
  when conservation is enforced (e.g., when squaring for probabilities),
that is associated with the volume of spacetime.  The new infinity can
thus be associated with conservation of the dilatation (scale) charge.  It
may be possible to absorb it into some renormalizations.

In loop diagrams the $x^0$ integration is more complicated, presumably
because of contributions corresponding to ``bulk modes".  Such
contributions should be compared to similar corrections coming from
higher orders in $R$.

\section{Dualities}

In the usual flat background there are two different (``dual") obvious
particle limits for strings:  $\alpha'\to 0$ and $\alpha'\to\infty$.  In both
these limits (expansions) one can still also expand in the string coupling
$g_s$.  The infinite tension ($\sim 1/\alpha'$) limit ($\alpha'\to 0$, or
``zero slope") describes the ground state of the string by shortening it to
a point (particle).  In the QCD string picture, this particle from the open
string is the $\rho$ meson (or other light mesons).  The flavor coupling
constant is identified with the open string coupling $\sqrt{g_s}$ (up to any
power of $\alpha'$ required to give it the right engineering dimensions).
(The limit $\alpha'\to 0$ can also give the classical mechanics limit of the
string, depending on how $\sigma$ is scaled.)

On the other hand, in the zero tension limit ($\alpha'\to\infty$) the
string ``falls apart":  There is no tension to hold the pieces together.  In
the QCD string picture, this is the limit where the gluons decouple.  Clearly
the tension is related to the color coupling constant.  Meanwhile, the
string coupling is identified with $1/N$ according to the 't Hooft
analysis, in terms of the number $N$ of colors.  Unfortunately, the usual
strings don't resemble QCD strings very closely, at least not when
expanded about flat space.  We now examine how these naive relations
are affected when expanding about $AdS_5\times S^5$.

We begin by analyzing how the two (dimensionless) coupling constants
appear in both the string and matrix field actions.  The worldsheet path
integral can consist of (1) the ``propagator" ($(\partial x)^2$) term, (2) a
Wess-Zumino term, (3) the (topological) worldsheet curvature term, (4)
a (worldsheet) cosmological constant term, and (5) the measure.
There is no Wess-Zumino term for the bosonic string; in any case, such a
term for superstrings introduces couplings only as higher-derivative
corrections to the leading interaction terms \cite{10}.

The curvature term is responsible for the topological $1/N$
coupling already considered:  It appears in the string action $S$,
contributing to the path integral as
$e^{-S}$, as a term
\be
S_\chi = \chi\ ln \left( {1\over N} \right) = \chi\, ln\,g_s
\quad\Rightarrow\quad g_s = {1\over N}
\ee
  where the Euler number $\chi$ is expressed as the integral of the
curvature.   Thus, in this approach the closed string coupling is simply
$1/N$ (and $1/\sqrt{N}$ for the open string), according to the general result
of 't Hooft.  This differs from the result of conventional random lattice
models because we make our analysis for the critical dimension, where
presumably there are no renormalization effects associated with a
worldsheet conformal anomaly.  (We have ignored the extra
compactification dimensions so far, but will return to them below, and
use the correct counting of dimensions for the D=10 superstring.)

The cosmological term is equivalent to the measure:
\be
\prod_I dx_I \; e^{\sum_I ln\; \mu_I} = \prod_I dx_I \; \mu_I
\ee
  Therefore, for convenience we associate such nonderivative terms with
the measure by definition.  Again assuming no renormalization effects
because of the critical dimensionality, the measure is then determined by
the same methods used in nonrelativstic quantum mechanics:  The exact
measure for path integrals can be derived by a lattice definition of path
integration in terms of ordinary integrals (as we do here) by starting
from a Hamiltonian form of the action, which follows directly from the
operator formalism.  Then there is no measure to start (other than the
usual $\sqrt{2\pi}$'s), but a measure is generated by integrating out the
momenta.  The result is a path integral of the form
\be
\prod_I \left( d^D x_I \; \sqrt{G}\, \right) \; exp\left[
	-\half \sum_{\langle IJ\rangle}
	G_{mn}(x^m_I -x^m_J)(x^n_I -x^n_J)\right]
\ee
  (where $G=det(G_{mn})$).  This is clearly covariant in the continuum limit,
since $d^D x \sqrt{G}$ is the usual covariant integration measure.  There
is some ambiguity in the discretized version (e.g., $G(x_I)$ vs.\
$G(x_J)$) related to higher-derivative corrections to the action; we have
made a choice consistent with the (worldsheet) continuum limit, and with
conformal invariance before the limit.  However, this expression for the
measure is exact including constant factors (once a $(2\pi)^{-D/2}$ is
included with each $d^D x$), which have no effect on covariance.  (For
example, the factor $(m/\hbar)^{D/2}$ appears for the free
nonrelativistic particle, with ``metric" $(dx)^2 m/\hbar$.)  These factors
are required to define the continuum limit:  For example, a spurious factor
of 2 at each vertex would generate a $2^V$ for $V$ vertices, which would
become infinite in the continuum limit $V\to\infty$.  In our case the
existence of such a limit is equivalent to worldsheet Weyl scale
invariance:  Integrating out half the vertices should produce a result of
the same form as the original, except for modification of the exponent
consistent with doubling the worldsheet area per vertex.  (The intrinsic 2D
``area" of the worldsheet lattice is the number of vertices.)

In our case the $R$ dependence for the bosonic string, before taking
the projective lightcone limit, is given by a measure factor of
$R$ accompanying the $dx^0$, following from the term $(Rdx^0/x^0)^2$ in
the string action.  Compactified dimensions (such as for
$S^5$) do not contribute:  Their measure must be normalized to give 1
upon integration, since in the projective lightcone limit they do not
contribute to the string action; i.e., $(\int d\Omega_5)^V=1^V\to 1$ as
$V\to\infty$.  ($R\to 0$ shrinks them to a point.  Before taking the limit
their measure is more complicated:  In the above derivation the momenta
are quantized; the sum is not as simple as a Gaussian integral, and
produces a more complicated measure factor.)

In the projective lightcone Feynman rules of the previous section, we can
then associate an $R$ with each vertex, with no extra factors from the
propagator (since the string action has no $R$ dependence in the
projective lightcone limit).  This corresponds to a matrix field Lagrangian
\be
L \to N\ tr\left(-\half\phi\Box\phi  -R{\textstyle{1\over 4}}\phi^4\right)
\ee
  (The coupling is negative so the vertex is positive:  The bosonic
functional integral is always positive, being the integral of an
exponential.)  We thus have the identifications
\be
g_s \;\sim\; {1\over N_c}\;, \qquad R \sim\; N_c g_c^2
\ee
  where $N_c$ is the rank of the ``color" gauge group of the gluon and
$g_c$ its coupling, while $g_s$ is the closed string (pomeron) coupling.
This result agrees with the naive qualitative relation between tension
($R^2$) and color coupling.

The Maldacena approach uses the other duality:  The particles of its
superconformal Yang-Mills theory are the ``massless" ground states of
the open superstring.  Thus the perturbation expansion used is different:
It corresponds neither to Regge nor parton behavior.  In that 
approach the identification between
(closed) string couplings and Yang-Mills ones is
\be
g_s \;\sim\; g_f^2 \; ,\quad R^4 \sim\; N_f g_f^2
\ee
  where $g_f$ is the open string (gauge meson) coupling, and $N_f$ is the
rank of the meson (``flavor") gauge group.  The resulting relations
\be
{1\over N_c} \;\sim\; g_f^2 \; , \quad N_c^4 g_c^8 \;\sim N_f g_f^2
\ee
  suggest a new type of duality between SYM theories.  (It is similar to
Seiberg's duality for $N=1$ theories \cite{16} in the sense that the rank
of the group is changed together with the coupling constant.)

\section{Superstring}

The classical $AdS_5\times S_5$ superstring action is \cite{17}
\be
S = \int d^2\sigma
\left[\half\sqrt{-g}g^{mn}L_m{}^{\cal A}(1) L_{n\cal A}(1)
+i\epsilon^{mn}\int_0^1ds\;
L_m{}^{\cal A}(s)\eta^{ij}\bar{\Theta}^i\gamma_{\cal A} L_n{}^j(s) \right]
\ee
where
\bea
d\sigma^m L_m{}^i(s) & = & \left(\frac{sinh (sM)}M
  D\Theta\right)^i \nonumber\\
d\sigma^m L_m{}^{\cal A}(s) & = & e_{\cal M}{}^{\cal A}(X)dX^{\cal M}
  -2i\bar{\Theta}^i\gamma^{\cal A}
  \left(\frac{sinh^2(\half sM)}{M^2}D\Theta \right)^i
\eea
and $M$ is a matrix bilinear in fermions, and $D\Theta$ a covariant
derivative on the fermions.  Here
$\eta^{ij}=\left({1\atop 0}{0\atop -1}\right)$ and
$\cal A,M$ are 10D vector indices.  In the (4D) covariant $\kappa$ gauge
\cite{18}
\be
\Theta^1 = \gamma_{0123}\Theta^2\quad (\gamma_{0123}=i\gamma_4)
\ee
  we obtain the gauge fixed action \cite{19,18} (in terms of the remainder
$\theta$ of
$\Theta$)
\bea
S = \int d^2\sigma \;\Bigg\{
\hskip-.25in&&  \half\sqrt{-g}g^{mn} \left[
{ \Pi_m{}^a \Pi_{ma} +R^2(\partial_m x^0)(\partial_n x^0)\over
(x^0)^2} +R^2(d\hat y)^2 \right] \nonumber \\[.1in]
&& -i\epsilon^{mn}{R(\partial_m\bar\theta)
   \gamma\cdot\hat y(\partial_n\theta)\over x^0}\Bigg\}
\eea
  in terms of the unit 6-vector $\hat y$ (of the internal SO(6)) on $S^5$ and
the 4D supersymmetric differential
\be
\Pi_m{}^a = \partial_m x^a -i\bar\theta\gamma^a\partial_m\theta
\ee
  where we have reinstated the dependence on $R$ as in section 2.
Remember that the usual AdS/CFT correspondence uses the JWKB
expansion.  Our new correspondence uses instead an expansion in $R$,
the leading term of which is the projective lightcone limit ($R\to 0$),
which  leaves only the first term,
\be
S = \int d^2\sigma\;\half\sqrt{-g}g^{mn}
{ \Pi_m{}^a \Pi_{ma}\over (x^0)^2}
\ee

The covariant gauge is more appropriate for expanding around ``long
string" configurations \cite{18}, for which one can use the static gauge
$x^0=\sigma^0$, $x^1=\sigma ^1$ to obtain a nondegenerate fermionic
kinetic operator ${\cal A}$ (${\cal A}^2\ne 0$).  (E.g., it was used in
\cite{20} to compute quantum corrections to the gauge theory $q\bar{q}$
potential via the AdS/CFT correspondence.)  For $\sigma$-independent
$x$-configurations (``short strings''), one can instead use the light-cone
gauge $(\gamma^3-\gamma^0)\Theta ^I=0$ \cite{21}, resulting in a more
complicated action (but with a nondegenerate fermionic kinetic operator).
Here we will ignore such subtleties; our approach can be applied for any
$\kappa$ gauge, and will suffer from the same difficulties, but we choose
the covariant gauge to illustrate how this covariance is reflected in the
form of the Feynman rules.

The projective lightcone superstring action is the same as the bosonic
one except for the supersymmetrization of the 4D differential
$\partial_m x^a\to\Pi_m{}^a$.  The discretized form is
\be
x_I^a-x_J^a \quad\to\quad
x_I^a-x_J^a-i\bar\theta_I\gamma^m\theta_J
\ee
  The Feynman rules then follow directly from the bosonic case:  As there,
$x^0$ dependence is determined by requiring that the propagator, vertex,
and external line factor be dimensionless.  (The dimensions of $\theta$
are $[\theta]=[x^{1/2}]$.)  The result for a $\phi^n$ interaction is then
\bea
\hbox{Propagator:} && d^{16}e^{-(x-x')^2/2x^0 x'^0}
   \delta^{16}(\theta -\theta ') \nonumber \\
\hbox{Vertex:} && \textstyle{\int} d^4 x\;d^{16}\theta\;dx^0 (x^0)^3
  \nonumber \\
\hbox{External line:} && (x^0)^{-4/n}\phi(x,\theta)
\eea
  Here we have used the identity
\be
f(x-x'-i\bar\theta\gamma\theta ') =
d^{16}f(x-x')\delta^{16}(\theta -\theta ')
\ee
for any function $f$, where $d$ is the covariant spinor derivative and
$d^{16}$ the antisymmetric product of all its components.  (This
can be derived, e.g., by writing $d^{16}=\int d^{16}\zeta\;exp(\zeta d)$.)

  Integrating out $x^0$ in tree graphs as before, the tree rules become
\bea
\hbox{Propagator:} && d^{16}
   \delta^{16}(\theta -\theta ')[\half (x-x')^2]^{4/n} \nonumber \\
\hbox{Vertex:} && \textstyle{\int} d^4 x\;d^{16}\theta\nonumber \\
\hbox{External line:} && \phi(x,\theta)
\eea
  (Certain values of $n$ give divergent numerical factors, which we
remove by ``renormalization".)  These imply the classical action
\be
S = N\ tr \int d^4 x\; d^{16}\theta\; (\half\phi d^{16}\Box^{4/n-6}\phi
+R{\textstyle{1\over n}}\phi^n)
\ee
  where we now include the coupling, which arises in the same way as
in the bosonic case.  Here we have used the identity
\be
  (d^{16})^2 \sim \Box^8 \quad\Rightarrow\quad
(d^{16})^{-1} \sim {d^{16}\over \Box^8}
\ee
  (This is easy to check in momentum space by choosing the lightcone
frame for the case $p^2=0$ and the rest frame otherwise.  In the former
case half the $d$'s have vanishing squares; in the latter the $d$'s are
proportional to $\gamma$ matrices for SO(16), and thus $d^{16}$ to
``$\gamma_5$".)

Unfortunately this action is nonlocal.  Its form is dictated by scale
invariance, once the interaction and $d^{16}$ numerator are assumed.
The nonlocality is probably due to the improper gauge fixing of $\kappa$
symmetry, which however preserves supersymmetry and scale
invariance.  We expect a better understanding of these issues in the
string action will lead to better particle field theory actions, as illustrated
by the bosonic $AdS_5$ string example, since the formalism (1) allows
explicit elimination of $x^0$, (2) produces propagators that are powers of
$x^2$ (or $p^2$), and (3) preserves the global symmetries of the string
action.  (Serious modifications are expected in order to avoid the no-go
theorem for maximally supersymmetric theories \cite{22}.)

For any covariant gauge, our approach produces manifestly $\cal N$=4
supersymmetric supergraphs and (gauge-fixed) actions.  Because of the
nonabelian gauge group, these would most likely be identified with the
lowest-spin $\cal N$=4 multiplet, super Yang-Mills.  After a clarification
of the above issues, this approach should give a better understanding of
the elusive $\cal N$=4 superspace formulation of this theory.

\section*{Acknowledgments}

This work was supported in part by NSF Grant PHY 9722101.
H.N. was supported in part by DoE Grant DE-FG02-90ER40542.
W.S. thanks Gordon Chalmers, Iouri Chepelev, Radu Roiban, and Diana
Vaman for helpful discussions.



\begin{thebibliography}{99}
\addtolength{\itemsep}{-1pt}

\bibitem{1} J. Maldacena, hep-th/9711200, {\it Adv. Theor. Math. Phys.}
{\bf 2} (1998) 231;\\
S.S. Gubser, I.R. Klebanov, and A.M. Polyakov, hep-th/9802109,
{\it Phys. Lett.} {\bf 428B} (1998) 105;\\
E. Witten, hep-th/9802150, {\it Adv. Theor. Math. Phys.} {\bf 2} (1998)
253;\\
O. Aharony, S.S. Gubser, J. Maldacena, H. Ooguri, and Y. Oz,
hep-th/9905111, {\it Phys. Rep.} {\bf 323} (2000) 183.

\bibitem{2} P.A.M. Dirac, {\it Ann. Math.} {\bf 37} (1936) 429;\\
H.A. Kastrup, {\it Phys. Rev.} {\bf 150} (1966) 1186;\\
G. Mack and A. Salam, {\it Ann. Phys.} {\bf 53} (1969) 174;\\
R. Marnelius and B. Nilsson, {\it Phys. Rev.} {\bf D22} (1980) 830.

\bibitem{3} S. Adler, {\it Phys. Rev.} {\bf D6} (1972) 3445.

\bibitem{4} M.F. Atiyah and R.S. Ward, {\it Comm. Math. Phys.} {\bf 55}
(1977) 117;\\
M.F. Atiyah, V.G. Drinfel'd, N.J. Hitchin, and Yu.I. Manin,
{\it Phys. Lett.} {\bf 65A} (1978) 185;\\
E. Corrigan, D. Fairlie, P. Goddard, and S. Templeton,
{\it Nucl. Phys.} {\bf B140} (1978) 31;\\
N.H. Christ, E.J. Weinberg, and N.K. Stanton, {\it Phys. Rev.} {\bf D18}
(1978) 2013;\\
M.F. Atiyah, {\it Geometry of Yang-Mills Fields} (Scuola Normale Superiore,
Pisa,  1979);\\
V.E. Korepin and S.L. Shatashvili, {\it Math. USSR Izvestiya} {\bf 24} (1985)
307.

\bibitem{5} R. Marnelius, {\it Phys. Rev.} {\bf D20} (1979) 2091;\\
W. Siegel, {\it Int. J. Mod. Phys. A} {\bf 3} (1988) 2713.

\bibitem{6} M.B. Halpern and W. Siegel, {\it Phys. Rev.} {\bf D16} (1977)
2486.

\bibitem{7} H.B. Nielsen and P. Olesen, {\it Phys. Lett.} {\bf 32B} (1970)
203;\\
D.B. Fairlie and H.B. Nielsen, {\it Nucl. Phys.} {\bf B20} (1970) 637;\\
B. Sakita and M.A. Virasoro, {\it Phys. Rev. Lett.} {\bf 24} (1970) 1146;\\
F. David, {\it Nucl. Phys.} {\bf B257 [FS14]} (1985) 543;\\
V.A. Kazakov, I.K. Kostov and A.A. Migdal, {\it Phys. Lett.} {\bf 157B} (1985)
295;\\
J. Ambj$\o$rn, B. Durhuus and J. Fr\"{o}hlich, {\it Nucl. Phys.} 
{\bf B257} (1985) 433.

\bibitem{8} M.R. Douglas and S.H. Shenker, {\it Nucl. Phys.} {\bf B335}
(1990) 635;\\
D.J. Gross and A.A. Migdal, {\it Phys. Rev. Lett.} {\bf 64} (1990) 127;\\
E. Br\'ezin and V.A. Kazakov, {\it Phys. Lett.} {\bf 236B} (1990) 144.

\bibitem{9} G. 't Hooft, {\it Nucl. Phys.} {\bf B72} (1974) 461.

\bibitem{10} A. Mikovic and W. Siegel, {\it Phys. Lett.} {\bf 240B} (1990)
363.

\bibitem{11} G. Veneziano, {\it Nuo. Cim.} {\bf 57A} (1968) 190;\\
V. Alessandrini, D. Amati, and B. Morel, {\it Nuo. Cim.} {\bf 7A} (1971) 797;\\
D.J. Gross and P.F. Mende, {\it Phys. Lett.} {\bf 197B} (1987) 129,
{\it Nucl. Phys.} {\bf B303} (1988) 407;\\
D.J. Gross and J.L. Ma\~nes, {\it Nucl. Phys.} {\bf B326} (1989) 73.

\bibitem{12} J.J. Atick and E. Witten, {\it Nucl. Phys.}
{\bf B310} (1988) 291.

\bibitem{13} W. Siegel, hep-th/9601002, {\it Int. J. Mod. Phys. A} {\bf 13}
(1998) 381.

\bibitem{14} G. Parisi and N. Sourlas, {\it Phys. Rev. Lett.} {\bf 43} (1979)
744.

\bibitem{15} W. Siegel, {\it Phys. Lett.} {\bf 142B} (1984) 276.

\bibitem{16} N. Seiberg, hep-th/9411149, {\it Nucl. Phys.} {\bf B435} (1995)
129.

\bibitem{17} R.R. Metsaev and A.A. Tseytlin, hep-th/9805028,
{\it Nucl. Phys.} {\bf B533} (1998) 109;\\
R. Kallosh and A.A. Tseytlin, hep-th/9808088, {\it JHEP} {\bf 9810} (1998)
016.

\bibitem{18} R. Kallosh and J. Rahmfeld, hep-th/9808038, {\it Phys. Lett.}
{\bf 443B} (1998) 143.

\bibitem{19} I. Pesando, hep-th/9808020, {\it JHEP} {\bf 9811} (1998) 002.

\bibitem{20} N. Drukker, D.J. Gross, and A. Tseytlin, hep-th/0001204,
{\it JHEP} {\bf 0004} (2000) 021.

\bibitem{21} I. Pesando, hep-th/9912284, {\it Phys. Lett.} {\bf 485B}
(2000) 246;\\
R.R. Metsaev and A.A. Tseytlin, hep-th/0007036.

\bibitem{22} W. Siegel and M. Ro\v cek, {\it Phys. Lett.} {\bf 105B} (1981)
275.

\end{thebibliography}
\end{document}